\documentclass[useAMS,usenatbib]{mn2e}
\usepackage{color,url,graphicx}
\usepackage{epstopdf}

\title[BL~Lac Black Hole Masses]{Dynamical Black Hole Masses of BL~Lac Objects from the Sloan Digital Sky Survey}
\author[Richard M. Plotkin et al.]{%
	R.~M.~Plotkin,$^{1}$\thanks{E-mail: r.m.plotkin@uva.nl}
	\ S.~Markoff,$^{1}$
	S.~C.~Trager,$^{2}$ 
	and S.~F.~Anderson$^{3}$ \\
	$^{1}$Astronomical Institute `Anton Pannekoek', University of Amsterdam, Science Park 904, 1098 XH, Amsterdam, the Netherlands
\\
	$^{2}$Kapteyn Astronomical Institute, University of Groningen, Postbus 800, NL-9700 AV Groningen, the Netherlands \\
	$^{3}$Department of Astronomy, University of Washington, Box 351580, Seattle, WA 98195, USA 
	}

\newcommand{\pasp}{PASP}
\newcommand{\aj}{AJ}
\newcommand{\mnras}{MNRAS}

\newcommand{\apj}{ApJ}
\newcommand{\apjl}{ApJL}
\newcommand{\araa}{AR\&A}
\newcommand{\aap}{A\&A}
\newcommand{\iaucirc}{IAU~Cric.}
\newcommand{\nat}{Nat}   
\newcommand{\bl}{BL~Lac}
\newcommand{\mbh}{M_{\rm BH}}
\newcommand{\msun}{{\rm M_{\sun}}}

\newcommand{\nhg}{143}                 
\newcommand{\nmbh}{71}               
\newcommand{\nnombh}{72}         
\newcommand{\nxray}{55}              
\newcommand{\ngalfit}{29}         
\newcommand{\nNogalfit}{42}    

\defcitealias{plotkin10}{P10}

\begin{document}

\date{}

\pagerange{\pageref{firstpage}--\pageref{lastpage}} \pubyear{2010}

\maketitle

\label{firstpage}


\begin{abstract}
We measure black hole masses for \nmbh\ \bl\ objects from the Sloan Digital Sky Survey (SDSS) with redshifts out to $z\sim0.4$.  We perform spectral decompositions of their nuclei from their host galaxies and measure their stellar velocity dispersions. Black hole masses are then derived from the black hole mass -- stellar velocity dispersion relation.   We find \bl\ objects host black holes of similar masses, $\sim$10$^{8.5}$~$\msun$, with a dispersion of $\pm$0.4~dex, similar to the uncertainties on each black hole measurement.   Therefore, all \bl\ objects in our sample have the same indistinguishable black hole mass.  These \nmbh\ \bl\ objects follow the black hole mass -- bulge luminosity relation, and their narrow range of host galaxy luminosities confirm previous claims that \bl\ host galaxies can be treated as standard candles.  We conclude that the observed diversity in the shapes of \bl\ object spectral energy distributions is not strongly driven by black hole mass or host galaxy properties.  
\end{abstract}



\section{Introduction}
\bl\ objects are a very rare class of Active Galactic Nuclei (AGNs) thought to be unified with Fanaroff-Riley~I \citep[FR~I,][]{fr74} radio galaxies viewed nearly along the axis of a relativistic jet \citep[e.g., see][]{blandford78,urry95}.   \bl\ objects are among the brightest radio, X-ray, and gamma-ray sources in the entire sky, they are highly polarized ($>$2--3\%), and their multiwavelength emission is variable on timescales ranging from minutes to decades \citep[e.g., see][]{kollgaard94,urry95}.    Perhaps the most striking property of \bl\ objects is their optical spectra, which are devoid of strong emission lines because radiation from the Doppler boosted jet is so dominant.   

Black hole mass measurements for a large number of \bl\ objects would have a wide range of applications -- from investigations aimed toward a better understanding of \bl\ statistical properties, to studies focusing more generally on relativistic jets.   For example, AGNs are powered by accretion onto a supermassive black hole, making black hole mass a fundamental parameter -- i.e., it sets an AGN's upper luminosity limit, and it determines the temperature of the accretion disc.  Also,  \bl\ spectral energy distributions (SEDs) are dominated by jet emission,  so they are  excellent probes of relativistic jet physics.   \bl\ objects show impressively diverse SEDs from object to object, with a synchrotron cutoff occurring anywhere from the near-infrared to the soft-X-ray wavebands \citep[e.g., see][]{padovani95_apj}.   A large sample of black hole masses can thus be used to investigate if  black hole mass or another parameter(s) is the primary driver of a jet's SED \citep[e.g., see][]{ghisellini08}.   Even more generally, a sample of \bl\ objects with well determined black hole masses could be incorporated in studies on the scaling of relativistic jets with mass, through, for example, comparison with X-ray binaries in the hard spectral state \citep[e.g., see][]{merloni03,falcke04}.  

 The weak-lined nature of \bl\ spectra, however, introduces unique challenges for obtaining reliable black hole mass measurements.  For other classes of AGN, black hole masses can be derived by measuring the widths of gas emission lines and by measuring the continuum luminosity -- here, one assumes the gas in the broad emission line region is virialized and adopts a relation between continuum luminosity and size of the broad emission line region \citep[e.g., see][]{kaspi00}.   However, this technique cannot be employed for AGN with featureless spectra.  Even if weak emission is detected (and if one assumes correlations between emission line strength and black hole mass extend to weak emission), the resultant \bl\ black hole masses will be unreliable unless the continuum luminosity is corrected for Doppler boosting.  Other techniques for weighing \bl\ central black holes should be seeked.

An alternative method utilizes the dynamics of the stars near a galaxy's center, i.e., `dynamical' mass estimates.  This is done  by measuring the line-of-sight stellar velocity dispersion ($\sigma$) from the widths of stellar absorption features, which are Doppler broadened due to the stars' motions.  Thus, if one can decompose an AGN's host galaxy from the central point source and measure $\sigma$, then black hole masses can be estimated from empirical correlations  \citep[the $\mbh$--$\sigma$ relation, e.g., see][]{ferrarese00,gebhardt00,tremaine02}.    Black hole masses can also be estimated from the host galaxy's luminosity  \citep[e.g.,][]{kormendy95,magorrian98},  and sizable samples of \bl\ black hole masses ($\sim$60-120 objects) have indeed been derived in such a way \citep[e.g.,][]{wu09, xu09}.    However,  the black hole mass -- galaxy luminosity correlation is not as tight as the $\mbh$--$\sigma$ relation, and black hole masses derived from galaxy luminosities can sometimes be overestimated by up to two orders of magnitude \citep[e.g., see][]{ferrarese00}.

Given the above considerations, we prefer to estimate black hole masses using stellar velocity dispersions.   However,  in the past, there was not  an abundant supply of \bl\ objects with high-quality spectra with enough apparent host galaxy light to perform the requisite measurements of $\sigma$.    For this reason, subsets of \bl\ objects with dynamical black hole estimates tend to contain at most $\sim$30 objects \citep[e.g., see][]{falomo03, woo05}.    Fortunately,  relatively large \bl\ samples containing hundreds of objects have recently come online \citep[e.g.,][]{turriziani07,massaro09}.    Here, we use the 723 object \bl\ sample assembled from the Sloan Digital Sky Survey \citep[SDSS,][]{york00} by \citet[][hereafter P10]{plotkin10}.  Each object has a high-quality SDSS spectrum, and the sheer size of this sample ensures a relatively large number of \bl\ spectra show a strong enough host galaxy component to measure $\sigma$.  We attempt black hole mass measurements for \nhg\ \citetalias{plotkin10} \bl\ objects, for which we derive \nmbh\ black hole masses.  \citet{leontavares10_ph} recently presented a similarly sized sample of dynamical black hole masses from our older radio-selected \bl\ catalog \citep{plotkin08}.   In \S \ref{sec:sample} we describe the parent \bl\ sample, and we explain our algorithm for estimating black hole masses in \S \ref{sec:bhmass}. The measured distribution of black hole masses and host galaxy luminosities are discussed in \S \ref{sec:disc}, and we comment on the import of black hole mass in determining the shape of \bl\ SEDs.    Finally, our conclusions are summarized in \S \ref{sec:conc}.  Throughout, we adopt $H_0=71$~km~s$^{-1}$~Mpc$^{-1}$, $\Omega_m=0.27$, and $\Omega_{\Lambda}=0.73$.

\section{The BL~Lac Sample}
\label{sec:sample}
The \citetalias{plotkin10} optically selected \bl\ sample contains 723 objects from 8250~deg$^2$ of SDSS spectroscopy.  About 75\% of these \bl\ objects were discovered by the SDSS.  In order for a source to be retained as a \bl\ candidate,  its SDSS spectrum cannot show any emission line with rest-frame equivalent width (REW) stronger than 5~\AA, and its Ca~{\tt II}~H/K depression must be smaller than 40\%.   There is no explicit requirement for an object to be a radio or X-ray emitter, but the sample was correlated post-selection with the NRAO VLA Sky Survey \citep[NVSS,][]{condon98} and with the Faint Images of the Radio Sky at Twenty-cm \citep[FIRST,][]{becker95} radio surveys, as well as with the {\it ROSAT} All Sky Survey \citep[RASS][]{voges99,voges00} in the X-ray.  We refer the reader to \citetalias{plotkin10} for details.  

About one-third of SDSS \bl\ objects have spectroscopic redshifts derived from host galaxy spectral features in their SDSS spectra.  Here, we limit ourselves to \nhg\ spectra that have reliable spectroscopic redshifts in \citetalias{plotkin10}, that match to a FIRST and/or NVSS radio source and are radio-loud (i.e., radio-to-optical flux ratio is larger than 10, see \citealt{kellermann89, stocke92}),\footnote{\citetalias{plotkin10} include a population of ~$\sim$10$^2$ weak-lined AGN lacking strong radio emission, many of which are unlikely best unified with \bl\ objects.  This constraint on radio emission therefore ensures we consider only normal \bl\ objects.}  
and that have $z<0.4$.   Objects with $z>0.4$ are too distant to show significant host galaxy radiation in their SDSS spectra.   We attempt to measure black hole masses for each of these \nhg\ \bl\ objects.   These \nhg\ objects are among the most weakly beamed \bl\ objects in \citetalias{plotkin10} (or else they would not show a host galaxy component in their optical spectra), thereby minimizing orientation effects.  
 
\section{Black Hole Mass Measurements}
\label{sec:bhmass}
To measure black hole masses we first remove the contribution from the AGN to the observed spectrum (\S \ref{sec:gandalf}), and we then measure the line of sight stellar velocity dispersion ($
\sigma$) in each decomposed host galaxy spectrum (\S \ref{sec:vdisp}).  We assess the reliability of the spectral decompositions in \S \ref{sec:badsp},  and in that section we identify  \nnombh\ spectra with unreliable decompositions (most have poor fits based on their reduced $\chi^2$).   Finally, we estimate black hole masses and host galaxy luminosities for the other \nmbh\ spectra  (\S \ref{sec:mbh} and \ref{sec:measHG}, respectively).

\subsection{Spectral Decomposition}
\label{sec:gandalf}
We assume each SDSS spectrum is a combination of two components:  the thermal contribution from the host galaxy, and the non-thermal emission from the relativistic jet.  We model the latter as a power law: $f_{\nu} = f_{\nu_0}\left(\nu/\nu_0\right)^{-\alpha_{\nu}}$, where  $\alpha_{\nu}$ is the spectral index and $f_{\nu_0}$ is the flux density at reference frequency $\nu_0$.  We choose $\nu_0$=c/(6165~\AA), which is the reference frequency of the SDSS $r$ filter.  To find the best fit power-law and host galaxy parameters, we make slight modifications to the Gas And Absorption Line Fitting algorithm ({\tt GANDALF}),\footnote{{\tt GANDALF} was developed by the SAURON team and is available from \url{http://www.strw.leidenuniv.nl/sauron}.  Also see \citet{sarzi06}.}  
 as described below.
 
{\tt GANDALF} iteratively fits an observed galaxy spectrum with combinations of stellar templates convolved by the best-fit line of sight velocity dispersion;  it also simultaneously fits a user defined list of gas emission lines, modeling the lines with Gaussian templates.    The Penalized Pixel-Fitting method (pPXF\footnote{{\tt IDL} code for running pPXF can be downloaded from  Michele Cappellari's web page: \url{http://www-astro.physics.ox.ac.uk/~mxc/idl/}.}, see \S \ref{sec:vdisp}) of \citet{cappellari04} is used by {\tt GANDALF} to  provide initial guesses for the stellar continuum fits and velocity dispersions.  Nonlinear  fits are then performed with the Bounded-Variables Least-Squares algorithm (BVLS\footnote{Code is also available from  Michele Cappellari's web page: \url{http://www-astro.physics.ox.ac.uk/~mxc/idl/}.}) and with Levenberg-Marquardt least-squares minimization \citep[see][]{markwardt09}.  An important feature of {\tt GANDALF} is that it fits the stellar continuum and gas emission simultaneously, instead of masking spectral regions containing emission lines from the continuum fits.  This technique is likely to yield less biased stellar continua \citep[see][]{sarzi06}.  

\begin{figure}
 \centering
\includegraphics[scale=0.42]{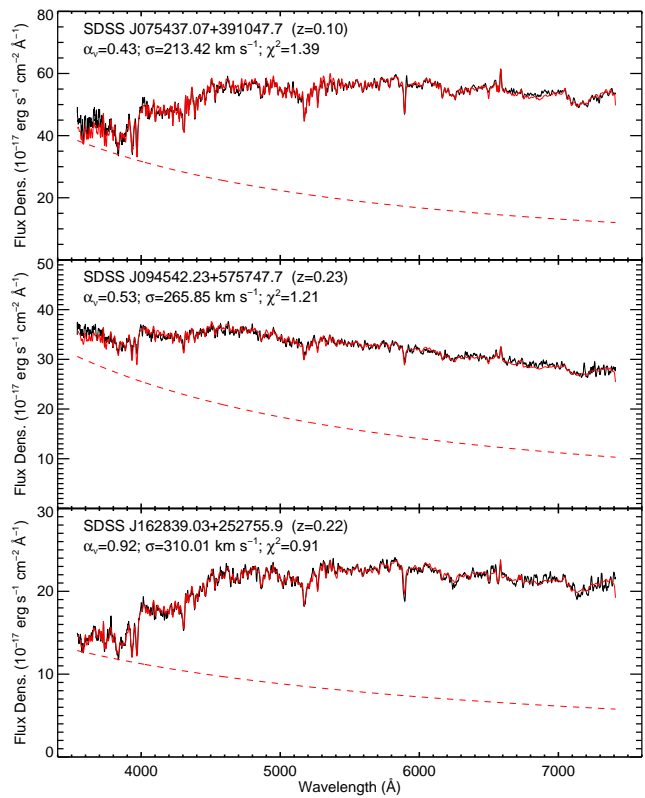}
 \caption{Three sample SDSS spectra, in their rest-frames, convolved to the MILES spectral resolution, shown as black solid lines.  The best {\tt GANDALF} fits are overdrawn as  red solid lines, and the red dashed lines illustrate the power law components.  This figure appears in color in the online version of this article.}
 \label{fig:sampspec}
 \end{figure}

We modified {\tt GANDALF} to also include a power law component from the AGN, and we then ran the modified code on the \nhg\ \bl\ objects showing a host galaxy component in  their SDSS spectra.   All fits are performed in the rest-frame.  The power law is constrained to always be positive and to not account for more than 80\% of the observed flux at 3900~\AA.  The latter constraint is to ensure the power law is physically meaningful: if the power law accounted for all of the observed flux, then we would not see any contribution from the host galaxy to the \bl\ spectrum.   We use stellar population models \citep{vazdekis10} based on the Medium-resolution Isaac Newton Telescope library of empirical spectra \citep[MILES,][]{sanchez06}, and the SDSS \bl\ spectra are convolved to match the spectral resolution of the MILES templates (81~km~s$^{-1}$).  Because \bl\ emission lines are by definition very weak, we only fit gas emission for [O~{\tt II}], [O~{\tt III}], H$\beta$, H$\alpha$, [N~{\tt II}], and [S~{\tt II}], and we exclude spectral regions containing Oxygen sky lines from the fits.    Sample spectral decompositions are shown in Fig.~\ref{fig:sampspec}.  

\subsection{Velocity Dispersion Measurements}
\label{sec:vdisp}
For each \bl\ spectrum, we subtract the best fit power law and emission lines, yielding a spectrum of just the host galaxy.  We then measure the host galaxy's velocity dispersion, $\sigma$, with the pPXF method.  The pPXF algorithm fits galaxy spectra by convolving sets of stellar templates with a best-fit velocity dispersion.   We again use the MILES stellar templates, and we adopt the best-fit combination of stellar templates found during the {\tt GANDALF} spectral decomposition in \S\ref{sec:gandalf}.    Since we remove any (weak) emission lines from our host galaxy spectra, there is no need to mask out regions of the spectra containing gas emission.  We do, however, still omit regions covering Oxygen sky lines from the pPXF fits.   

We estimate errors on $\sigma$ via Monte Carlo simulations \citep[see][]{cappellari04}.   For each pixel's flux density, $f_{\lambda,i}$, we add noise $\pm \epsilon_i$, where $\epsilon_i$ is drawn randomly from a normal distribution with standard deviation equal to each pixel's flux density measurement error.  We then remeasure the velocity dispersion with pPXF, repeating 100 times.  We adopt the standard deviation in the 100 velocity dispersion measurements as the formal uncertainty on $\sigma$, which is typically better than 10\%.   Uncertainties on power law indices are similarly derived by randomly adding noise and re-running {\tt GANDALF} 100 times per spectrum.   The best-fit stellar velocity dispersions and their uncertainties are included in Table~\ref{tab:bhmass}. 

 In Table~\ref{tab:bhmass} we also include the best fit power law indices, $\alpha_{\nu}$, and the fraction of the observed flux density accounted for by the power law at rest-frame 6165~\AA, $f_{pl}$.  The measured values of $\alpha_{\nu}$ are typical for \bl\ objects, with an average  $\left<\alpha_{\nu}\right>=0.70\pm0.56$.    The recovered values of 
 $f_{pl}$ (with an average value $0.34\pm0.12$) are as expected.  If the power law is too weak it would not be a \bl\ object, and if the power law is too strong the host galaxy would not be seen in the optical spectrum.  The fraction of light accounted for by the power law is systematically higher if the calculation is instead done at a wavelength bluer than 6165~\AA.

\begin{table*}
\begin{minipage}{150mm}
\caption{Black Hole Masses and Other Measured Parameters}
\label{tab:bhmass}
\scriptsize
\begin{tabular}{c rr c rc c cc cc cc}
\hline
		Name      &  
	         RA \ \           &  
	         	Dec \ \ 	      &   
		Redshift  &  
		$\alpha_{\nu}$ & 
		$\sigma_{\alpha_{\nu}}$ & 
		$f_{pl}$$^a$ &
		$\sigma_{\star}$ & 
		$\sigma_{\sigma_{\star}}$ & 
		$\log \mbh$    & 
		 $\sigma_{\mbh}$ & 
		 $M_{hg}$    & 
		 $\sigma_{M_{hg}}$ \\ 
		 (SDSS J)   & 
		 (deg.)         & 
		 (deg.)         & 
		                     & 
		                      & 
		                      &
		                      & 
		   (km~s$^{-1}$) & 
		   (km~s$^{-1}$) & 
		  ($\msun$)    & 
		  (dex)    & 
		  (mag)         & 
		  (mag)         \\ 
\hline
002200.95$+$000657.9 &      5.50396 &      0.11610 &    0.306 &     0.55 &     0.21 &     0.16 &      246 &       13 &     8.49 &     0.32 &   -22.42 &     0.14 \\
005620.07$-$093629.7 &     14.08366 &     -9.60826 &    0.103 &     0.80 &     0.05 &     0.33 &      332 &        9 &     9.01 &     0.32 &   -23.26 &     0.30 \\
075437.07$+$391047.7 &    118.65446 &     39.17994 &    0.096 &     0.43 &     0.28 &     0.29 &      213 &        5 &     8.24 &     0.31 &   -22.95 &     0.08 \\
080018.79$+$164557.1 &    120.07831 &     16.76588 &    0.309 &     1.40 &     0.15 &     0.47 &      259 &       13 &     8.58 &     0.32 &   -23.17 &     0.32 \\
082323.24$+$152447.9 &    125.84686 &     15.41333 &    0.167 &     0.50 &     0.14 &     0.23 &      293 &       12 &     8.80 &     0.32 &   -22.76 &     0.05 \\
082814.20$+$415351.9 &    127.05917 &     41.89776 &    0.226 &     1.12 &     0.04 &     0.27 &      299 &        9 &     8.83 &     0.32 &   -23.04 &     0.30 \\
083417.58$+$182501.6 &    128.57328 &     18.41712 &    0.336 &     0.54 &     0.35 &     0.18 &      399 &       37 &     9.34 &     0.36 &   -22.87 &     0.32 \\
083548.14$+$151717.0 &    128.95059 &     15.28808 &    0.168 &     1.14 &     0.14 &     0.53 &      179 &       14 &     7.94 &     0.33 &   -22.76 &     0.31 \\
083918.74$+$361856.1 &    129.82812 &     36.31559 &    0.335 &     0.61 &     0.11 &     0.20 &      247 &       27 &     8.50 &     0.36 &   -22.61 &     0.34 \\
084712.93$+$113350.2 &    131.80388 &     11.56396 &    0.198 &    -0.22 &     0.19 &     0.35 &      251 &       15 &     8.52 &     0.32 &   -22.87 &     0.09 \\
085036.20$+$345522.6 &    132.65085 &     34.92296 &    0.145 &     0.76 &     0.11 &     0.33 &      263 &        8 &     8.61 &     0.31 &   -23.19 &     0.05 \\
085729.78$+$062725.0 &    134.37411 &      6.45695 &    0.338 &    -0.45 &     0.71 &     0.09 &      212 &       21 &     8.23 &     0.35 &   -22.70 &     0.36 \\
085749.80$+$013530.3 &    134.45751 &      1.59176 &    0.281 &     0.54 &     0.17 &     0.37 &      276 &       13 &     8.69 &     0.32 &   -23.91 &     0.10 \\
090207.95$+$454433.0 &    135.53316 &     45.74250 &    0.289 &     1.07 &     0.01 &     0.48 &      291 &       11 &     8.78 &     0.32 &   -22.43 &     0.33 \\
090314.70$+$405559.8 &    135.81128 &     40.93330 &    0.188 &     0.37 &     0.09 &     0.20 &      218 &        8 &     8.28 &     0.31 &   -23.09 &     0.07 \\
090953.28$+$310603.1 &    137.47201 &     31.10088 &    0.272 &    -0.56 &     0.41 &     0.21 &      321 &       16 &     8.95 &     0.32 &   -23.78 &     0.31 \\
091045.30$+$254812.8 &    137.68876 &     25.80358 &    0.384 &     1.30 &     0.02 &     0.41 &      249 &       31 &     8.51 &     0.38 &   -22.82 &     0.36 \\
091651.94$+$523828.3 &    139.21642 &     52.64121 &    0.190 &     0.66 &     0.05 &     0.34 &      252 &        7 &     8.53 &     0.31 &   -23.68 &     0.07 \\
093037.57$+$495025.6 &    142.65655 &     49.84045 &    0.187 &     0.41 &     0.15 &     0.41 &      244 &       20 &     8.48 &     0.34 &   -22.64 &     0.31 \\
094022.44$+$614826.1 &    145.09354 &     61.80727 &    0.211 &     0.85 &     0.09 &     0.33 &      258 &       13 &     8.57 &     0.32 &   -22.48 &     0.08 \\
094542.23$+$575747.7 &    146.42599 &     57.96325 &    0.229 &     0.53 &     0.09 &     0.44 &      266 &       14 &     8.63 &     0.32 &   -23.39 &     0.30 \\
101244.30$+$422957.0 &    153.18461 &     42.49918 &    0.365 &     0.47 &     0.35 &     0.47 &      272 &       25 &     8.67 &     0.35 &   -23.03 &     0.28 \\
102453.63$+$233234.0 &    156.22348 &     23.54278 &    0.165 &     1.06 &     0.04 &     0.51 &      137 &        9 &     7.46 &     0.33 &   -22.83 &     0.30 \\
102523.04$+$040228.9 &    156.34603 &      4.04139 &    0.208 &     1.23 &     0.08 &     0.44 &      206 &       16 &     8.18 &     0.33 &   -22.26 &     0.31 \\
103317.94$+$422236.3 &    158.32478 &     42.37678 &    0.211 &     0.75 &     0.08 &     0.20 &      260 &        8 &     8.59 &     0.31 &   -22.91 &     0.05 \\
104029.01$+$094754.2 &    160.12090 &      9.79839 &    0.304 &     0.42 &     0.01 &     0.35 &      277 &       13 &     8.70 &     0.32 &   -23.05 &     0.31 \\
104149.15$+$390119.5 &    160.45480 &     39.02209 &    0.208 &     1.01 &     0.10 &     0.22 &      255 &       11 &     8.55 &     0.32 &   -23.41 &     0.06 \\
104255.44$+$151314.9 &    160.73103 &     15.22083 &    0.307 &     1.23 &     0.17 &     0.55 &      166 &       33 &     7.81 &     0.46 &   -22.16 &     0.35 \\
105344.12$+$492955.9 &    163.43387 &     49.49889 &    0.140 &     1.31 &     0.04 &     0.42 &      244 &        9 &     8.47 &     0.31 &   -23.13 &     0.05 \\
105538.62$+$305251.0 &    163.91095 &     30.88085 &    0.243 &     0.81 &     0.12 &     0.20 &      238 &       17 &     8.43 &     0.33 &   -22.72 &     0.31 \\
105606.61$+$025213.4 &    164.02756 &      2.87041 &    0.236 &     1.13 &     0.13 &     0.37 &      197 &       11 &     8.11 &     0.32 &   -22.58 &     0.31 \\
105723.09$+$230318.7 &    164.34625 &     23.05522 &    0.378 &     0.83 &     0.17 &     0.37 &      223 &       19 &     8.32 &     0.34 &   -23.17 &     0.24 \\
112059.74$+$014456.9 &    170.24892 &      1.74914 &    0.368 &     0.14 &     0.21 &     0.20 &      465 &       18 &     9.60 &     0.33 &   -22.96 &     0.34 \\
113630.09$+$673704.3 &    174.12539 &     67.61789 &    0.134 &     1.14 &     0.04 &     0.44 &      221 &        7 &     8.30 &     0.31 &   -22.65 &     0.05 \\
114023.48$+$152809.7 &    175.09784 &     15.46937 &    0.244 &    -0.38 &     0.14 &     0.30 &      428 &       32 &     9.46 &     0.35 &   -23.96 &     0.09 \\
114535.10$-$034001.4 &    176.39625 &     -3.66708 &    0.168 &    -0.48 &     0.29 &     0.14 &      216 &        9 &     8.27 &     0.31 &   -22.78 &     0.07 \\
115404.55$-$001009.8 &    178.51899 &     -0.16941 &    0.254 &     0.45 &     0.07 &     0.37 &      228 &       16 &     8.36 &     0.33 &   -22.53 &     0.31 \\
115709.53$+$282200.7 &    179.28974 &     28.36687 &    0.300 &     0.56 &     0.14 &     0.26 &      369 &       21 &     9.20 &     0.33 &   -23.52 &     0.10 \\
120837.27$+$115937.9 &    182.15533 &     11.99387 &    0.369 &     0.66 &     0.14 &     0.45 &      271 &       22 &     8.66 &     0.34 &   -23.35 &     0.34 \\
123123.90$+$142124.4 &    187.84962 &     14.35680 &    0.256 &     1.64 &     0.03 &     0.71 &      264 &       33 &     8.62 &     0.38 &   -22.72 &     0.32 \\
123131.39$+$641418.2 &    187.88081 &     64.23841 &    0.163 &    -0.16 &     0.05 &     0.25 &      300 &        8 &     8.84 &     0.31 &   -23.37 &     0.30 \\
123831.24$+$540651.8 &    189.63018 &     54.11441 &    0.224 &     1.07 &     0.07 &     0.30 &      263 &       12 &     8.61 &     0.32 &   -23.09 &     0.06 \\
125300.95$+$382625.7 &    193.25398 &     38.44049 &    0.371 &     1.09 &     0.41 &     0.41 &      214 &        7 &     8.24 &     0.31 &   -22.36 &     0.37 \\
131330.12$+$020105.9 &    198.37552 &      2.01832 &    0.356 &     0.98 &     0.01 &     0.45 &      248 &       28 &     8.50 &     0.36 &   -22.96 &     0.36 \\
132231.46$+$134429.8 &    200.63112 &     13.74164 &    0.377 &     0.51 &     0.29 &     0.28 &      323 &       17 &     8.97 &     0.33 &   -23.25 &     0.33 \\
132239.31$+$494336.2 &    200.66380 &     49.72673 &    0.332 &     0.26 &     0.26 &     0.15 &      273 &       24 &     8.67 &     0.35 &   -23.00 &     0.19 \\
132301.00$+$043951.3 &    200.75419 &      4.66427 &    0.224 &     0.37 &     0.12 &     0.22 &      304 &       14 &     8.86 &     0.32 &   -23.17 &     0.30 \\
132617.70$+$122958.7 &    201.57378 &     12.49965 &    0.204 &     0.30 &     0.11 &     0.25 &      266 &       13 &     8.63 &     0.32 &   -22.87 &     0.31 \\
133612.16$+$231958.0 &    204.05067 &     23.33280 &    0.267 &     0.30 &     0.19 &     0.31 &      256 &       16 &     8.56 &     0.33 &   -22.80 &     0.32 \\
134105.10$+$395945.4 &    205.27127 &     39.99595 &    0.172 &     1.52 &     0.03 &     0.50 &      244 &       13 &     8.48 &     0.32 &   -22.43 &     0.31 \\
134136.23$+$551437.9 &    205.40096 &     55.24388 &    0.207 &     0.86 &     0.06 &     0.38 &      219 &       15 &     8.29 &     0.33 &   -22.71 &     0.31 \\
134633.98$+$244058.4 &    206.64162 &     24.68291 &    0.167 &     0.81 &     0.05 &     0.32 &      219 &       11 &     8.29 &     0.32 &   -22.29 &     0.06 \\
135314.08$+$374113.9 &    208.30870 &     37.68721 &    0.216 &     1.11 &     0.05 &     0.36 &      291 &       10 &     8.79 &     0.32 &   -23.64 &     0.10 \\
140350.28$+$243304.8 &    210.95951 &     24.55133 &    0.343 &     1.76 &     0.45 &     0.57 &      233 &       37 &     8.39 &     0.41 &   -22.22 &     0.39 \\
142421.17$+$370552.8 &    216.08824 &     37.09802 &    0.290 &     2.38 &     0.89 &     0.53 &      233 &       11 &     8.39 &     0.32 &   -23.21 &     0.22 \\
142832.60$+$424021.0 &    217.13587 &     42.67252 &    0.129 &    -0.21 &     0.14 &     0.30 &      277 &       11 &     8.70 &     0.32 &   -23.19 &     0.05 \\
144248.28$+$120040.2 &    220.70118 &     12.01119 &    0.163 &    -0.20 &     0.14 &     0.31 &      319 &       15 &     8.94 &     0.32 &   -23.27 &     0.30 \\
144932.70$+$274621.6 &    222.38626 &     27.77269 &    0.227 &     0.66 &     0.09 &     0.33 &      303 &       18 &     8.86 &     0.33 &   -22.85 &     0.31 \\
153311.25$+$185429.1 &    233.29688 &     18.90809 &    0.307 &     0.59 &     0.18 &     0.44 &      312 &       13 &     8.91 &     0.32 &   -23.01 &     0.32 \\
155412.07$+$241426.6 &    238.55032 &     24.24073 &    0.301 &     0.05 &     0.31 &     0.21 &      260 &       16 &     8.59 &     0.33 &   -22.81 &     0.16 \\
155424.12$+$201125.4 &    238.60054 &     20.19040 &    0.222 &     0.50 &     0.11 &     0.29 &      317 &        9 &     8.94 &     0.32 &   -23.48 &     0.05 \\
160118.96$+$063136.0 &    240.32900 &      6.52667 &    0.358 &     1.06 &     0.01 &     0.41 &      276 &       11 &     8.69 &     0.32 &   -23.06 &     0.33 \\
160519.04$+$542059.9 &    241.32937 &     54.34998 &    0.212 &     0.67 &     0.11 &     0.33 &      171 &       18 &     7.85 &     0.36 &   -21.86 &     0.32 \\
161541.21$+$471111.7 &    243.92174 &     47.18661 &    0.199 &     1.31 &     0.03 &     0.45 &      204 &       10 &     8.17 &     0.32 &   -22.72 &     0.30 \\
161706.32$+$410647.0 &    244.27636 &     41.11307 &    0.267 &     1.76 &     0.03 &     0.61 &      169 &       11 &     7.84 &     0.33 &   -22.79 &     0.31 \\
162839.03$+$252755.9 &    247.16263 &     25.46553 &    0.220 &     0.92 &     0.06 &     0.32 &      310 &        9 &     8.90 &     0.32 &   -23.06 &     0.05 \\
163726.66$+$454749.0 &    249.36112 &     45.79695 &    0.192 &     1.27 &     0.07 &     0.29 &      237 &       10 &     8.42 &     0.31 &   -22.44 &     0.05 \\
164419.97$+$454644.3 &    251.08321 &     45.77900 &    0.225 &     0.85 &     0.06 &     0.32 &      288 &       13 &     8.76 &     0.32 &   -23.06 &     0.30 \\
205456.85$+$001537.7 &    313.73690 &      0.26049 &    0.151 &     0.15 &     0.32 &     0.23 &      272 &        8 &     8.67 &     0.31 &   -22.63 &     0.22 \\
205938.57$-$003756.0 &    314.91071 &     -0.63222 &    0.335 &     0.68 &     0.01 &     0.31 &      114 &        2 &     7.16 &     0.32 &   -22.84 &     0.32 \\
223301.11$+$133602.0 &    338.25465 &     13.60056 &    0.214 &     0.27 &     0.26 &     0.29 &      253 &       14 &     8.54 &     0.32 &   -22.67 &     0.31 \\
\hline
\footnotetext[1]{Ratio of the flux density from the power law to the total observed flux density at rest-frame 6165~\AA.}
\end{tabular}
\end{minipage}
\end{table*}

\subsection{Assessing the Quality of the Decomposition}
\label{sec:badsp} 
There are several reasons why some of the spectral decompositions might fail.  All \nhg\ SDSS \bl\ spectra are of high enough quality to derive redshifts from host galaxy spectral features, and to allow reliable \bl\ classifications.  However, the amount of the host galaxy flux for the most highly beamed objects in this subset is not strong enough to obtain accurate spectral decompositions, and some other objects do not have high enough signal-to-noise.  We thus exclude 54 objects with reduced $\chi^2_r>2$ (each {\tt GANDLAF}  fit has $\sim$4200 degrees of freedom), and we remove an additional 12  objects for which the spectral decompositions did not converge, or the error on the velocity dispersion is larger than 40~km~s$^{-1}$ (i.e., half the spectral resolution of the MILES stellar templates).  Upon visual examination of each spectral fit, we remove an additional six objects with poor fits.  Most of these six objects have Ca~{\tt II}~H/K break strengths smaller than 15\%, and their continua are too dominated by the power law to obtain reliable $\sigma$ measurements.  We are left with \nmbh\ spectra for which we determine black hole masses.    For the remainder of this paper we only consider these \nmbh\ objects.

The \nmbh\ successfully decomposed objects are clearly biased toward those with strong host galaxy components.   The amount of stellar flux in \bl\ spectra depends primarily on orientation.  That is, more weakly beamed \bl\ objects show more stellar light \citep[e.g.,][]{landt02, plotkin08}, and our \nmbh\ object subset is therefore biased toward larger viewing angles.  All of the 71 \bl\ objects with successful decompositions have Ca~{\tt II}~H/K breaks stronger than 15\%, roughly translating to viewing angles larger than about 15-25 deg \citep[see Fig.~8 of][]{landt02}. Since the above bias is primarily geometric, we make the reasonable assumption that our analysis and conclusions on this subset can be extrapolated to the entire \bl\ population.

As pointed out by \citet{woo05}, less massive black holes live in less luminous galaxies and are therefore more difficult to measure since the host galaxy is harder to see, especially at increasing redshift.  This selection effect could artificially narrow the observed stellar velocity (and therefore black hole mass) distribution.  We however do not see any correlation between velocity dispersion and redshift in our sample of \nmbh\ objects, and there is no obvious deficit of low mass black holes at high-redshift.  We therefore do not believe this strongly affects our sample.  Regardless, since we do not attempt to derive a complete sample of \bl\ objects with black hole masses here, this potential bias will not alter our conclusions.   

\subsection{Measuring Black Hole Masses}
\label{sec:mbh} 
\begin{figure}
 \centering
\includegraphics[scale=0.43]{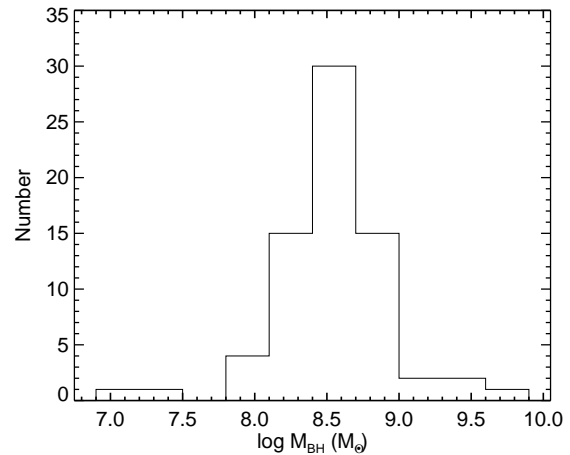}
 \caption{Distribution of black hole mass measurements, derived with Equation~\ref{eq:msigma}.}
 \label{fig:mbhdist}
 \end{figure}

The $M_{\rm BH}$--$\sigma$ relation is parameterized as 
\begin{equation}
\log \left(\mbh/\msun \right) = \alpha + \beta \log \left(\sigma/\sigma_0 \right),
\label{eq:msigma}
\end{equation}
with $\sigma_0=200$~km~s$^{-1}$.    We use the coefficients derived by \citet{tremaine02} from  a sample of 31 nearby galaxies: $\alpha=8.13\pm0.06$ and $\beta=4.02\pm0.32$.   To estimate error bars, we propagate the uncertainties on our velocity dispersion measurements and the errors on $\alpha$ and $\beta$.  We add that uncertainty in quadrature with the $\mbh$--$\sigma$ relation's intrinsic scatter in $\log \mbh$: $\sigma_{\rm int} \sim \pm 0.30$~dex \citep[see][]{tremaine02}.  Typical error bars on our black hole measurements, including the intrinsic scatter of the $\mbh-\sigma$ relation,  are $\pm$0.30--0.35~dex.  The measured black hole mass distribution is shown in Fig~\ref{fig:mbhdist}, and masses are listed along with the other fitted parameters (i.e., power law indices, velocity dispersions, etc.)  in Table~\ref{tab:bhmass}.

There appears to be a small number of black hole mass measurements that deviate from the mean black hole mass in Figure~\ref{fig:mbhdist} (i.e., $\log \mbh < 7.5$ and $\log \mbh > 9.0$).   We re-examined their spectral decompositions visually, and we find no compelling reason to remove them from the sample.  Instead, we attribute those points to populating the high and low sigma tails of the mass distribution, and to small number statistics.  A Kolmogorov-Smirnoff test shows our measured black hole mass distribution is consistent with a Gaussian, in which case we would expect 3.6 black hole mass measurements to deviate from the mean by $\pm$2$\sigma$.  Indeed, four black hole mass measurements populate the 2$\sigma$ tail.

\subsection{Measuring Host Galaxy Luminosities}
\label{sec:measHG}
After performing the spectral decomposition we are able to construct spectra of each \bl\ object's host galaxy.  We use these spectra to estimate the absolute magnitude each host galaxy would have in the Cousins~$R_c$ filter at $z=0$.  First, we measure the host galaxy flux at rest-frame 6581~\AA\ (the effective wavelength of the $R_c$~filter), and we then apply a bandpass correction and correct for extinction from the Milky Way (using the dust maps of \citealt{schlegel98}).  

The 3$''$ apertures of each SDSS spectroscopic fiber does not include all of the host galaxy's flux, and an aperture correction must be applied.  To determine the fraction of light covered by each spectroscopic fiber, we perform an image decomposition on each source's sdss $i$ filter image using the {\tt GALFIT} software \citep{peng02}.  We model each \bl\ object with a point source,\footnote{The point-spread function at the location of each \bl\ object on the SDSS camera is extracted with the {\tt read\_PSF} code provided by the SDSS collaboration: \url{http://www.sdss.org/dr7/products/images/read_psf.html}.} 
 and a de Vaucouleur profile (\bl\ objects probably live exclusively in elliptical galaxies, \citealt{urry00_hstii}).  We are able to measure the half light effective radius $r_e$ for \ngalfit\ \bl\ hosts, requiring $r_e>3$~pix, and that the best-fit host galaxy and central point source models are centered on the same coordinates.  We adopt these $r_e$ values to calculate the aperture correction for these \ngalfit\ objects.  For the remaining objects, we assume $r_e=10$~kpc, the median effective radius for the \ngalfit\ objects. This value is also consistent with the median effective radius from other \bl\ host galaxy studies \citep[e.g.,][]{urry00_hstii}.

The above aperture correction contributes the largest source of uncertainty to our luminosity estimates.  For the \ngalfit\ objects with $r_e$ measurements, the uncertainties on $r_e$ are typically of the order $\pm$1~kpc (0.6 pix), introducing typical errors on the host galaxy absolute magnitudes of $\pm$0.06~mag.   The systematic error introduced for the other \nNogalfit\ objects by assuming $r_e=10$~kpc is more severe.  To estimate the uncertainty, we calculate the host galaxy luminosity for the \ngalfit\ objects with $r_e$ measurements by assuming $r_e=10$~kpc, and we compare to the host galaxy luminosity estimates using each $r_e$ measurement.  There is no systematic offset between the two absolute magnitude measurements; however, the measurements disagree with a standard deviation of $\pm$0.3~mag.  We therefore adopt a systematic uncertainty of $\pm$0.3~mag for the \nNogalfit\ objects without $r_e$ measurements.  Note, the error's dependence on redshift is weak for $z>0.1$.

We add the above uncertainties in quadrature to the statistical errors estimated from Monte Carlo simulations, where we randomly added noise to each spectrum and re-ran each spectral decomposition 100 times  (as described in \S \ref{sec:vdisp}).    These uncertainties, including error from the aperture corrections, are listed  in Table~\ref{tab:bhmass}.  For the \ngalfit\ objects with $r_e$ measurements, we do not estimate absolute magnitudes using the total integrated host galaxy apparent magnitude from their best-fit de Vaucouleur profiles because the required K-corrections  would introduce additional uncertainties.

\section{Discussion}
\label{sec:disc}
\subsection{Black Hole Masses}
\label{sec:disc_mbh}

We measure an average black hole mass of $\left < \log \mbh \right> = 8.54 \pm 0.047~\msun$ (the quoted uncertainty is the error of the mean), with a standard deviation of $\pm$0.40~dex.  The dispersion in the distribution of black hole masses is comparable to the typical measurement error of $\pm$0.30--0.35~dex.  Thus, it is possible that the observed spread of black hole masses can be attributed only to measurement error, and we conclude \bl\ black hole masses are virtually indistinguishable.

Our black hole estimates are consistent with those in the literature.  For example, \citet{falomo03} find $\left < \log \mbh \right> =  8.57~\msun$ for a sample 12 \bl\ objects with dynamical black hole mass measures.   \citet{woo05} find 32 \bl\ objects with dynamical black hole mass measures to have a dispersion from $4\times10^7 - 6\times10^8~\msun$, and $\left < \log \mbh \right> \sim 8.30~\msun$.  They note, however, that their sample is relatively shallow  ($\left<z\right>\sim0.17$), and it thus may not probe a large enough volume to recover a substantial number of massive central black holes.  Indeed, when they include an additional $\sim$30 \bl\ objects with black hole mass measures based on host galaxy luminosity, their $\sim$60 object sample has $\left<z\right>\sim0.31$ and a dispersion in black hole mass measures that extends to $4\times10^9~\msun$ (identical to the largest black hole mass measured in our \nmbh\ object sample). 

It is interesting that the black hole mass distribution is so narrow, because \bl\ SEDs can be incredibly diverse: \bl\ objects are often classified as high-energy peaked (HBL) or low-energy peaked (LBL) \bl\ objects, depending on the cutoff frequency of their synchrotron radiation \citep[see][]{padovani95_apj}.    \citet{nieppola06} show \bl\ synchrotron radiation cutoff frequencies can span close to ten orders of magnitude from object to object.    Yet, we find that differences in SED shapes must not significantly depend on black hole mass.

 \begin{figure}
 \centering
\includegraphics[scale=0.5]{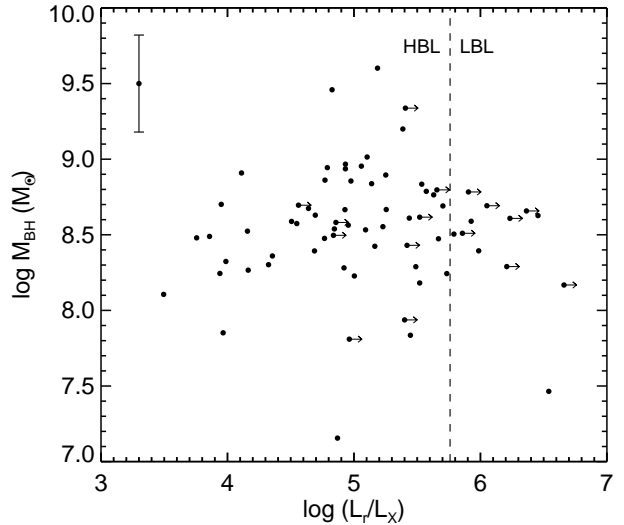}
 \caption{Logarithm of black hole mass vs.\ logarithm of the ratio of radio (5~GHz) to X-ray (1~keV) specific luminosities.   The latter parameterizes the shape of each object's SED.   Error bars are omitted for clarity --  a typical (1$\sigma$) black hole mass measurement error is shown in the top left corner.  The measurement errors on $\log\left(L_{r}/L_{X}\right)$ are smaller than the data points.  We do not observe a significant correlation indicating one or more parameters other than black hole mass are important for determining the SED shape.}
 \label{fig:sed}
 \end{figure}

We parameterize the SED shape of our \nmbh\ \bl\ objects with $\log \left(L_{r}/L_{X}\right)$, where $L_{r}$ and $L_{X}$ are the specific luminosities (in erg~s$^{-1}$~Hz$^{-1}$) at rest-frames 5~GHz and 1~keV respectively, and their values are taken from \citetalias{plotkin10}.  The SEDs of \bl\ objects with smaller $L_r/L_X$ values have synchrotron cutoffs at higher frequencies.  We follow \citet{plotkin08} and classify objects with $\log \left(L_{r}/L_{X}\right)<5.76$ as HBLs.  Our sample of \nmbh\ \bl\ objects includes 59 HBLs and 12 LBLs;  the relatively larger number of HBLs in our sample is an artifact of SDSS \bl\ selection \citepalias[see][]{plotkin10}.  A Kolmogorov-Smirnoff test indicates a p=0.47 chance that the HBL and LBL black hole mass distributions come from the same population, and their black hole mass distributions are therefore not statistically different. 

The logarithm of black hole mass vs.\ $\log \left(L_r/L_X\right)$ is shown in Fig.~\ref{fig:sed}, and we see no statistically significant correlation.  The linear Pearson correlation coefficient between $\log \left(L_r/L_x\right)$ and $\log \mbh$ is $r=0.018$, with a probability $p=0.44$ of randomly finding a stronger correlation, treating $L_r/L_X$ limits as exact.  We find $r=0.036$ and $p=0.40$ if we only consider the \nxray\ objects with X-ray detections in RASS.   Thus, differences between \bl\ SEDs are likely not driven solely by black hole mass.  This is consistent with \citet{ghisellini08}, who show the observed range of \bl\ synchrotron cutoff frequencies can be replicated by assuming all \bl\ objects have the same black hole masses and varying only their Eddington normalized accretion rates (see their Fig.~5).  We do not derive normalized accretion rates for the \nmbh\ \bl\ objects with black hole mass measurements here because, without knowing their Doppler factors, we cannot accurately debeam their luminosities.

\subsection{Host Galaxies}
\label{sec:hg}

\begin{figure}
 \centering
\includegraphics[scale=0.5]{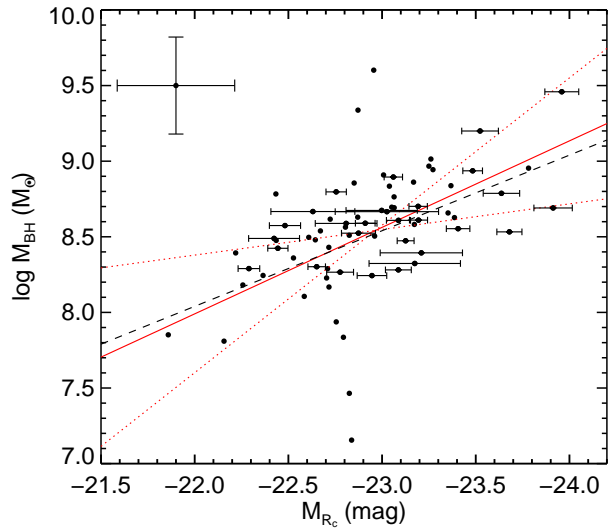}
 \caption{Logarithm of black hole mass vs.\ host galaxy absolute magnitude in the Cousins~$R_c$ filter.  Error bars on absolute magnitude are shown for the \ngalfit\ objects with effective radius measurements.  For clarity, a typical error bar for the other \nNogalfit\ objects (where we assume $r_e=10$~kpc) is shown in the top left corner, as is the typical $1\sigma$ uncertainty for all \nmbh\  black hole mass measurements.  The solid red line shows our best-fit linear regression using the Bayesian method of \citet{kelly07}, and the red dotted lines illustrate the $\pm$3$\sigma$ uncertainties.  The black hole mass -- bulge relationship from \citet{mclure02} is overplotted as a black dashed line for comparison. This figure appears in color in the online version of this article.}
 \label{fig:bhbulge}
 \end{figure}

A correlation between black hole mass and host galaxy absolute magnitude ($M_{R_c}$) is found (Fig.~\ref{fig:bhbulge}), confirming that \bl\ objects follow the black hole mass -- bulge relationship \citep[e.g.,][]{kormendy95}.  Our sample contains the largest number of \bl\ objects for which the correlation has been tested.     We perform a linear regression using the technique of \citet{kelly07}, which takes a Bayesian approach and accurately accounts for measurement errors in both variables.  We find $\log \left (M_{BH}/M_{\sun}\right) = \left(-4.12\pm2.99\right) - \left(0.55 \pm 0.13\right) M_{R_c}$, and a linear Pearson's correlation coefficient $r=-0.513$ ($p<10^{-5}$ chance of randomly finding a stronger correlation).    Our regression agrees with the correlation found by \citet{mclure02} for a sample of 90 active and inactive galaxies covering similar redshift ranges and host galaxy luminosities (dashed line in Fig.~\ref{fig:bhbulge}).  

There are a small number of points that deviate from the black hole mass - bulge relation in Fig.~\ref{fig:bhbulge} (i.e., those with $\log \mbh < 7.5$ and $\log \mbh > 9.0$).  There is no motivation to remove those objects from our sample based on the quality of their spectral decompositions, and we thus include all objects in the above regression.  We however note they merit further study.

There is no reason to assume the observed correlation is driven by correlated errors between our \bl\ black hole mass and host galaxy luminosity measurements.   We test this by randomly adding noise to each observed spectrum and rerunning our spectral decomposition 100 times for each \bl\ object.  No source shows a significant correlation between its 100 pairs of $\log \mbh$ and $M_{R_c}$: the average Pearson's correlation coefficient for the \nmbh\ sources is $\left<r\right>=-0.01$, and no source shows an anti-correlation stronger than $r=-0.2$.   

  That \bl\ objects obey the same black hole mass -- bulge relation as other galaxies indicates that the black hole formation histories of \bl\ objects are similar to other AGN.  This is further support that  parameters other than black hole mass (such as orientation, accretion rate, etc.), must be important for dictating if an AGN looks like a \bl\ phenomenon, and then for determining if the \bl\ object is an LBL or HBL.  

\begin{figure}
 \centering
\includegraphics[scale=0.5]{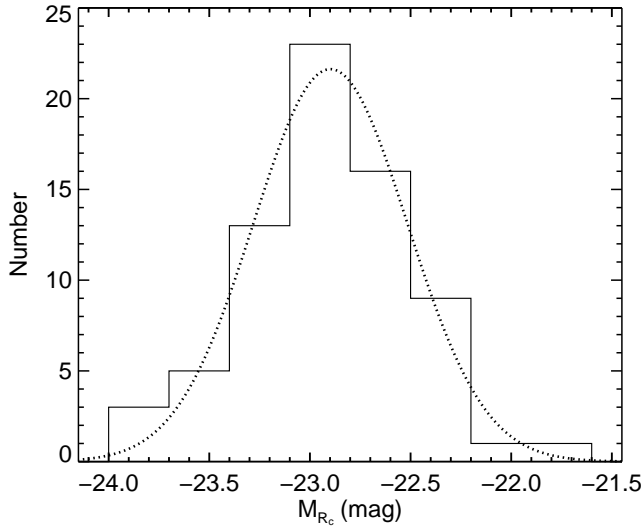}
 \caption{Absolute magnitude of each decomposed host galaxy in the $R_c$ filter.  The distribution of absolute magnitudes is well described by a Gaussian with mean -22.9~mag and standard deviation $\pm$0.4~mag.  This confirms earlier claims that the luminosities of \bl\ host galaxies is narrow enough to treat them as standard candles.  We do not see any differences in the host luminosities of LBLs or HBLs, indicating the host galaxy does not play a significant role in determining their SEDs.}
 \label{fig:absmag}
 \end{figure}

With the black hole mass -- bulge relation in mind,  we expect a small scatter in the \bl\ host galaxy luminosity distribution, since they have a relatively narrow range of black hole masses.  This is indeed observed, as shown in Fig.~\ref{fig:absmag}.  The \bl\ host galaxies are well fit by a Gaussian with $\left< M_{R_c} \right >=-22.9$~mag and a standard deviation of just $\pm$0.4~mag.  This confirms the ``standard candle'' result of \citet{sbarufatti05}, who showed 64 \bl\ objects with their host galaxies imaged by the {\it Hubble} Space Telescope have host galaxy absolute magnitudes well-fit by a Gaussian with mean $-22.8$~mag and standard deviation $\pm$0.5~mag (using the same cosmology employed herein).    For the redshifts considered here, the distribution is sufficiently narrow to estimate redshifts better than $\Delta z \sim \pm 0.07$--0.08 (e.g., see \citealt{sbarufatti05}, \citealt{meisner10}, \citetalias{plotkin10}).  \citet{meisner10} and \citetalias{plotkin10} apply the standard candle result to show that \bl\ objects lacking spectroscopic redshifts tend to be more distant on average, which can have important implications for understanding \bl\ evolution.
 
 We find no significant difference between the distribution of host galaxy luminosities for HBLs and LBLs.  The same host galaxy absolute magnitude ($\left< M_{R_c}\right> = -22.9$~mag) is measured for both HBLs and LBLs, and the Student's t-test indicates a probability $p=0.977$ that both distributions have the same mean assuming their variances are similar ($p=0.973$ if we allow for different variances).  The Student's t-test assumes the data follow a normal distribution, which is justified by our observations (see Fig.~\ref{fig:absmag}).  Even if we drop this assumption,  the non-parametric Mann-Whitney U-Test still indicates  the sample means are not significantly different (at the $p=0.379$ level).  Finally, a Kolmogorov-Smirnoff test gives a $p=0.820$ chance the samples are drawn from the same parent population.  We thus conclude that the observed diversity in \bl\ SEDs is not caused by differences in host galaxy properties.

\section{Conclusions}
\label{sec:conc}
AGN very likely influenced the formation of galaxies and heirarchical structure in the universe \citep[e.g.,][]{dimatteo05}; obtaining reliable AGN black hole mass estimates  is therefore an important endeavor.  We presented dynamical black hole mass measurements for \nmbh\ \bl\ objects that show enough host galaxy contribution to their SDSS spectra to measure their stellar velocity dispersions.  

   We find SDSS \bl\ objects have similar black hole masses, with the dispersion in their distribution comparable to the measurement uncertainty ($\sim$0.30--0.35~dex).  We do not see any trend between black hole mass and the ratio of radio to X-ray luminosity, which serves as a proxy for SED shape.  Our conclusion is that the diverse SED shapes exhibited by \bl\ objects is not controlled solely by black hole mass; another parameter, perhaps accretion rate, must be important \citep[e.g., see][]{ghisellini08}. Dynamical black hole mass measures can only be obtained for the most weakly beamed \bl\ objects, or else their optical spectra do not show a strong enough host galaxy component.  Therefore, the \nmbh\ \bl\ objects with black hole measures have similar orientation angles, and relativistic beaming is likely secondary in determining the shape of their SEDs.  However, orientation effects may not be completely negligible \citep[e.g., see][]{nieppola08}.

We also recover a correlation between black hole mass and host galaxy luminosity that is consistent with the black hole mass -- bulge luminosity relation in the literature \citep[e.g.,][]{mclure02}.  This indicates the formation histories of \bl\ central black holes is similar to other AGN; we do not see any evidence that SED shape depends on host galaxy properties either.   Our host galaxy decomposition also supports claims that the distribution of \bl\ host galaxy luminosities is narrow enough to treat them as standard candles.  One can thus derive ``imaging redshifts'' for \bl\ objects lacking spectroscopic redshifts if their host galaxies can be resolved from the nucleus, and redshift lower limits can be estimated when host galaxies are not detected.

The black hole masses presented here expand the types of applications of SDSS \bl\ catalogs.  For example, relativistic outflows are observed not just from supermassive black holes but also from their much smaller Galactic ($\sim$10$~\msun$) counterparts.  The discovery of the fundamental plane of black hole accretion, a correlation between radio luminosity, X-ray luminosity, and black hole mass, demonstrates a remarkable connection between accreting black holes over eight orders of magnitude of black hole mass \citep[e.g.,][]{merloni03,falcke04}.    Part of our motivation for measuring \bl\ black hole masses is to investigate how well \bl\ objects fit onto the fundamental plane; this will be discussed in a future paper.   \bl\ objects are an interesting AGN subclass for fundamental plane studies because their non-thermal emission is dominated by the jet.  

\section*{Acknowledgments}
We thank the anonymous referee for insightful comments that improved this manuscript.  R.M.P. and S.M. acknowledge support from a Netherlands Organization for Scientific Research (NWO) Vidi Fellowship.  S.M. also acknowledges support from The European CommunityÕs Seventh Framework Programme (FP7/2007-2013) under grant agreement number ITN 215212 ÒBlack Hole Universe.Ó

\bibliographystyle{mn2e}


\bsp

\label{lastpage}

\end{document}